# Numerical Analysis of Cavitation Dynamics on Free Ogee Spillways Using the Volume of Fluid (VOF) Method

Parvaneh Nikrou[a] , Sajjad Pirboudaghi[b]

[a] *Department of Geography, The University of Alabama, Tuscaloosa, AL 35487, USA*
*pnikrou@crimson.ua.edu*

[b] *Assistant professor, Engineering faculty of Khoy, Urmia university of technology, Urmia, Iran*
*S.pirboudaghi@uut.ac.ir*

## Abstract

Simulating complex hydraulic conditions, particularly two-phase flows over spillway chutes, can be achieved with high accuracy using three-dimensional numerical models. This study investigates the potential for vacuum generation and cavitation phenomena on the Aghchai Dam service spillway through numerical simulations conducted in Flow-3D. The analysis focuses on two specific flow rates, 4400 and 1065 cubic meters per second, as determined by experimental data. The Volume of Fluid (VOF) method is employed to accurately calculate the free surface flow. Simulation results at a discharge rate of 4400 cubic meters per second indicate a high likelihood of cavitation at critical locations, including the ogee curve and the angle transition in the chute channel. These areas require specific mitigation measures to prevent cavitation-induced damage. In contrast, at the lower flow rate of 1065 cubic meters per second, the risk of cavitation is minimal due to reduced flow velocity and the absence of flow separation from the bed. The numerical findings align closely with empirical observations, demonstrating the reliability of the simulation approach in predicting cavitation behavior.

## 1. Introduction

Spillways are critical hydraulic structures designed to safely convey excess water from reservoirs. However, these structures are highly susceptible to cavitation, a phenomenon that can cause significant damage to their surfaces. Cavitation occurs when flow lines separate from the spillway floor due to surface irregularities, leading to localized pressure drops. If the pressure falls below the fluid's vapor pressure, the liquid begins to vaporize, forming bubbles. These bubbles are transported downstream to areas of higher pressure, where they collapse violently, generating shock waves that can erode and corrode the solid surfaces of the spillway. The pressure caused by bubble collapse can exceed 1000 megapascals, creating immense localized forces on the spillway bed and stilling basins. Repeated cavitation events lead to cumulative damage, resulting in large cavities known as cavitation erosion.

Given the significant risks associated with cavitation, it is imperative to adopt robust monitoring and risk mitigation and management strategies(1-3). Incorporating advanced technologies, such as artificial intelligence (4)and machine learning (5), combined with unmanned aerial vehicles (UAVs)(6-8), offers an effective solution for structural inspection(9) and hazard identification. These methods enable the early detection of cavitation damage, enhance predictive maintenance, and optimize resource allocation. The application of explainable machine learning frameworks in complex physical systems—including recent advances in speech modeling using Mel-Frequency Cepstral Coefficients (MFCCs) for capturing acoustic variation—demonstrates how computational methods can provide both interpretability and precision across disciplines (10). At the same time, the communication of uncertainty and technical evaluation in engineering domains often requires careful rhetorical framing. Recent corpus-based research has shown that even in academic writing within empirical fields, hedging is commonly used to present findings with measured caution and mitigate criticism (11). Additionally, public education (12) on hazard risks(13), coupled with improved project management(14) for maintenance, is crucial. This includes providing appropriate access routes(15-19) for inspections in hazardous areas and considering the environmental implications of potential spillway failures. Such comprehensive strategies not only safeguard spillway functionality but also promote sustainable and resilient dam management practices. Moreover, remote sensing technologies and flood inundation mapping(20) have significantly improved our ability to predict hydraulic risks, offering accurate assessments of spillway performance and aiding in emergency planning(25-21).

Spillways are also influenced by seepage and structural irregularities, which can exacerbate cavitation risks. Seepage analysis using finite element methods, as demonstrated in studies like those on the Sahand rockfill dam, highlights the importance of optimal structural configurations to reduce hydraulic stresses and enhance spillway resilience. Implementing measures such as clay blankets and concrete covers has been shown to reduce discharge rates and prevent seepage-induced damage(21, 26-28). These findings underscore the need for continuous improvement in spillway design and maintenance practices(29).

Even minor surface irregularities, such as abrupt changes in cross-section or protrusions caused by construction or operational wear, can initiate flow line separation and cavitation. These irregularities are often found at critical points, such as the bases of valves, beneath sliding gates, or near the transition regions of chutes. For instance, spillway flows occurring 40–50 meters below the reservoir water level are particularly prone to cavitation.

Over the decades, extensive experimental, numerical, and analytical research has been conducted to understand cavitation and develop mitigation strategies for spillways. Model-prototype relationships have been extensively studied to improve the reliability of hydraulic models. Experimental investigations by Momber(30) revealed that materials with higher toughness exhibit better resistance to cavitation damage by distributing local stresses more effectively. Similarly, Samadi et al. (31) demonstrated that increased surface and bed roughness reduces cavitation intensity. On stepped spillways, Frizell et al. (32) observed significant cavitation effects, with Matos et al. (33) highlighting the role of flow rate and step height in increasing cavitation risks.

Flow aeration has been extensively studied as an effective method to mitigate cavitation. Chanson (34) explored aeration techniques, while Felder and Chanson (35) demonstrated that porosity in stepped spillways reduces cavitation likelihood by enhancing aeration. Other studies, such as those by Dong et al. (36) emphasize the importance of optimal aeration location and geometry in minimizing cavitation risks. Advances in numerical methods have further enabled accurate modeling of turbulent, two-phase flows over spillways using approaches like the Reynolds-averaged Navier–Stokes (RANS) equations, the Renormalization Group (RNG) turbulence model, and the Volume of Fluid (VOF) method.

Cavitation-induced damage remains a critical challenge for hydraulic structures operating at high velocities, particularly in high dams and large spillways. Addressing this persistent issue requires comprehensive risk assessment and the implementation of effective mitigation strategies. The AghChai Dam, recognized as the nation's first arched rockfill dam, holds significant importance in regional agriculture and drinking water supply. Given the substantial investment in its construction, optimizing operational efficiency and reducing maintenance costs are essential.

One of the primary concerns is the mitigation of cavitation-induced surface erosion to enhance the mechanical efficiency and longevity of the structure. This study leverages advanced numerical modeling to evaluate the cavitation potential and flow characteristics of the AghChai Dam spillway. Using Flow-3D software, the research simulates various flow regimes to estimate the onset and extent of cavitation while analyzing the spillway's outlet discharge. The outcomes of this investigation aim to minimize erosion-related expenses and ensure the spillway's reliable performance under high-flow conditions.

This paper is structured to provide a investigation into the occurrence of cavitation on free ogee spillways, with a particular focus on the Aghchai Dam spillway. Following the introduction, a review of the relevant literature highlights the key principles of cavitation, its impact on hydraulic structures, and prior research employing both experimental and numerical approaches to study this phenomenon. The methodology section details the geometric and hydraulic characteristics of the Aghchai Dam spillway, alongside the use of Flow-3D software (37), the Volume of Fluid (VOF) method, and turbulence models for simulating the complex two-phase flow conditions. The results section presents findings from the simulations, comparing cavitation potential and discharge behavior under varying flow conditions, and validating the numerical outcomes against empirical data. Finally, the discussion and conclusion offer insights into the implications of the findings, proposing strategies for mitigating cavitation risks and enhancing the efficiency of spillway designs

## 2. Research Methodology

In this section, the methodology employed to investigate the occurrence of cavitation phenomena on free ogee spillways is presented. The study utilizes the Reynolds-averaged Navier–Stokes (RANS) equations as the governing equations for simulating turbulent flows, incorporating both continuity and momentum equations to describe the flow field. The Volume of Fluid (VOF) method is applied to model the free-surface behavior of the fluid, while turbulence effects are captured using an eddy viscosity approach. The Flow-3D software is selected for its ability to simulate complex hydraulic conditions, including cavitation, turbulence, and flow interactions with solid boundaries. The software's numerical framework, based on finite volume and difference approximations, enables precise modeling of both fluid and structural components using the FAVOR method. This section provides a comprehensive overview of the mathematical framework, numerical techniques, and simulation setup used to achieve accurate predictions of cavitation conditions under varying discharge rates.

## 2.1. Governing Equations for Turbulent Flow

The averaged Navier-Stokes equations, often known as Reynolds (RANS)(38), represent the equations governing the movement of a viscous incompressible fluid in a turbulent state.

- **Continuity:**

The flow continuity equation is determined by using the law of mass conservation and calculating the mass balance equation for a fluid element. This equation is generally written as:

$$V_F \frac{\partial \rho}{\partial t} + \frac{\partial}{\partial x}(\rho u A_x) + R\frac{\partial}{\partial y}(\rho v A_y) + \frac{\partial}{\partial z}(\rho w A_z) + \xi \frac{\rho u A_x}{x} = 0 \quad (1)$$

where (VF) is the ratio of the fluid volume passing through an element to the total volume of the element and (ρ) is the fluid density. The velocity components (u,v,w) are in (x,y,z) directions. (Ax) is the ratio of the area of the fluid traveling through an element to the total area of the element in the (x) direction, whereas (Az) and (Ay) are similarly the ratio of the flow levels in the (z) and (y) directions. (R) and (ξ) are related to the type of coordinate system in Cartesian coordinates, R=1 and ξ=1.

- **Momentum:**

The momentum equation in the x direction is calculated as follows. Obviously, it can be used for other directions.

$$\frac{\partial u_i}{\partial t} + \frac{\partial u_i u_j}{\partial x_j} = -\frac{1}{\rho}\frac{\partial P}{\partial x_i} + g_i + \frac{\partial}{\partial x_j}(\tau_{ij}) \quad (2)$$

In the above equations, $u_i$ is the velocity component in the $x_i$ direction, P is the total pressure, ρ is the fluid density, $g_i$ is the acceleration of gravity in the $x_i$ direction, and $\tau_{ij}$ is the stress tensor, which is expressed as the following equation in the case of turbulent flow:

$$\tau_{ij} = \left[ \rho(\nu + \nu_t)\left( \frac{\partial u_i}{\partial x_j} + \frac{\partial u_j}{\partial x_i} \right) \right] - \frac{2}{3}\rho(k + \nu_t)\frac{\partial u_i}{\partial x_i}\delta_{ij} \tag{3}$$

In turbulent flows, shear stress is composed of two terms: in addition to the shear stress caused by the average component of the flow, another shear stress induced by the fluctuating components of the velocity is known as Reynolds stress and is represented by the following equation:

$$\tau_{ij} = -\rho\overline{u_i' u_j'} = \rho \nu_t \left(\frac{\partial u_i}{\partial x_j} + \frac{\partial u_j}{\partial x_i}\right) - \frac{2}{3}\rho k \delta_{ij} \tag{4}$$

In the above equations, $v_t$ is the eddy viscosity or turbulence viscosity, which unlike molecular viscosity is not a fluid characteristic. It depends on the characteristics of the flow and its turbulence, and its value varies from fluid to fluid and from point to point. δij (Kronecker's delta) is for applying the eddy viscosity model (EVM). The kinetic energy of turbulence in the unit of mass (k) is expressed as follows:

$$k = \frac{1}{2}\left(\overline{u_i'}^2 + \overline{u_j'}^2 + \overline{u_k'}^2\right)$$

$$\begin{cases} \delta_{ij} = 0 & i \neq j \\ \delta_{ij} = 1 & i = j \end{cases} \tag{5}$$

To solve the turbulent flow field employing continuity and Reynolds equations, the Reynolds stresses in the equations must be modelled in a certain way. The presence of four equations (1 continuity and three momentum) and four unknowns in the flow field determines the 3-dimensional flow in this scenario (that is, the velocities in 3 directions x, y and z, plus pressure).

Reynolds stresses are expressed using turbulence models. Eddy viscosity is presented using various theories (such as mixing theory) in the aforementioned models, and the relationship between Reynolds stresses and average velocity components is established. It is obvious that numerical methods are required to solve the following equations. Computational fluid dynamics (CFD) refers to mathematical models that use numerical approaches to solve fluid fields. CFD simulates system behavior such as fluid flow, heat transport, and other related processes. Flow-3D is a sophisticated software in the field of computational fluid dynamics that is produced, developed, and supported by Flow Science, Inc. It covers many physical models such as cavitation, turbulence, determining the flow pattern on different types of spillways, and flow on breakwaters. This software includes two numerical techniques: the volume of fluid method (VOF)(39) for displaying fluid behavior on a free surface and the Fractional Area/Volume Obstacle Representation (FAVOR) (40) for simulating solid surfaces and volumes. Flow-3D uses finite volume and difference approximations to solve the governing equations of fluid movement.

## 2.2. Governing Equation of the Volume of Fluid (VOF) Method

The VOF transfer method in Flow-3D is implemented using the donor-acceptor cell approximation. This standard approach employs a splitting operator and the VOF function's value to calculate the flux of fluid across all three coordinate directions. On a free surface, not all computational elements contain fluid uniformly. As illustrated in Figure 1, certain elements are completely filled, while others are entirely void, and a subset of elements are partially filled, representing the free surface. These partially filled elements define the interface between the fluid and the surrounding void, enabling precise tracking of the free surface dynamics.

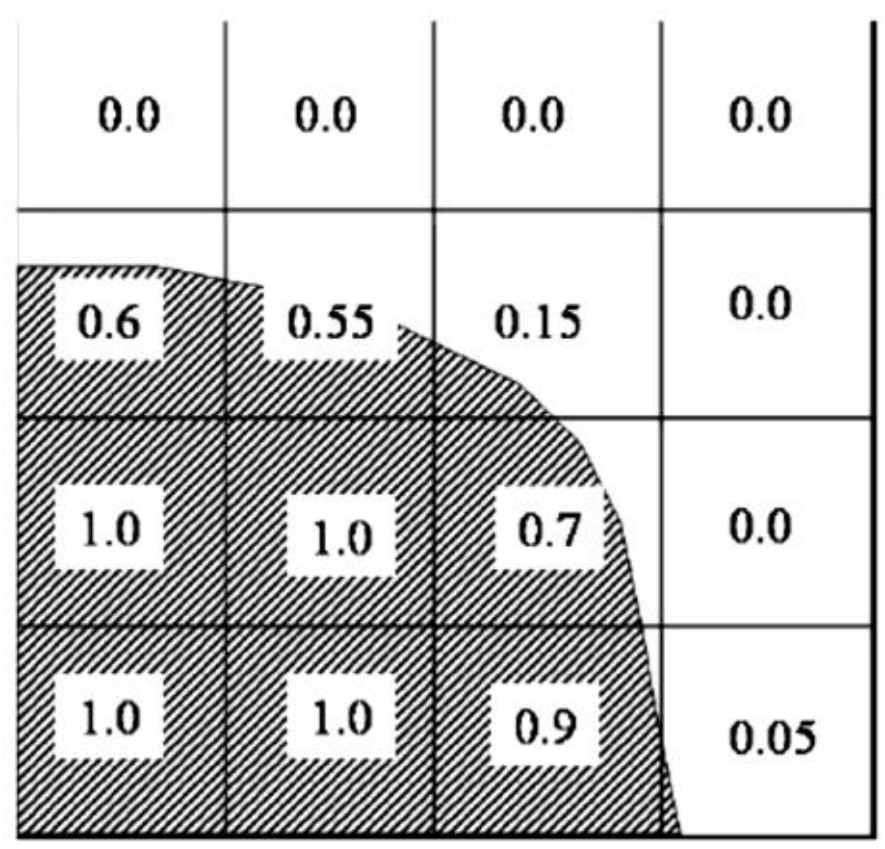

*Figure 1 An example of the values of the VOF function near the free surface*

The fluid fraction function (F), commonly referred to as the volume of fluid, takes a value of 1 within the fluid domain and 0 outside of it. The position and orientation of the free surface within a computational cell can be determined by the value of F. This positioning is influenced by the fluid distribution in adjacent cells, as the fluid within the surface elements tends to align with neighboring elements containing the highest fluid volume. When the value of F for a control volume lies between 0.0 and 1.0, it signifies the presence of a free surface within the element. In two dimensions, the differential form of this function is expressed as follows:

$$\frac{\partial F}{\partial t} + u\frac{\partial F}{\partial x} + v\frac{\partial F}{\partial y} = 0 \tag{6}$$

In this equation, the value of F is 1 for a fully fluid-filled cell and 0 for a cell completely devoid of fluid. For surface cells, where the fluid interface is present, F takes a value between 0 and 1, representing the fraction of the cell volume occupied by fluid. The viscosity and density parameters for each cell are determined using the Navier-Stokes equations in conjunction with the VOF equation. The interface between the two fluid phases is captured by their combined contributions to the density and viscosity in the surface cells, reflecting the local mixture of the phases.

$$\rho = \sum_{i=1}^{n} F_i \rho_i \qquad \mu = \sum_{i=1}^{n} F_i \mu_i \tag{7}$$

The fluid configuration is represented by the VOF, $F(x, y, z, t)$, which varies depending on the type of problem being modeled. For incompressible flow scenarios, such as a fluid with a free surface or two fluids sharing an interface, F represents the fraction of a cell volume occupied by the fluid. A value of F=1 indicates the presence of fluid, while F=0 corresponds to voids or bubbles. These voids and bubbles are regions devoid of fluid mass, where a uniform pressure is maintained. Physically, such spaces are typically filled with vapor or gas, whose density is negligible compared to that of the liquid phase.

## 2.3. Technical Specifications of the Dam

The Aghchai River flows through the Chaipara Plain, where the Aghchai Dam is situated 42 kilometers from Khoy City and 35 kilometers from Chaipara City (Figure 2). The irrigation and drainage network associated with the dam serves the agricultural lands of the Chaipara and Nazak Plains. To safely manage floodwaters entering the Aghchai Dam reservoir, a free spillway was constructed on the dam's left side (Figure 3) (41). This spillway channels flood discharge into a waterway on the dam's left side, eventually directing it back into the main river. The spillway and flood drainage system are designed to handle the maximum probable flood, ensuring that water levels remain below the dam's crest, even during a 10,000-year flood combined with a wind event with a 10-year return period. The design considers wave height and surge effects to prevent overtopping.

The spillway system adheres to USBR standards and includes an ogee crest profile, chute walls, and a stilling basin. These elements are optimized to control negative pressures and ensure sufficient water depth without the need for additional freeboard. The system also features a directing channel at the upstream end to guide flow towards the control section. A trapezoidal channel, excavated in front of the control section, has a base elevation of 1292.5 meters above sea level. Its width transitions from 115.5 meters at the upstream end to 80 meters at the control section. The channel side slopes are 1:1, and the length along the overflow axis is approximately 78.5 meters. The maximum flow velocity is designed to be 3 meters per second.

The spillway's ogee-shaped control section spans 80 meters in length. The ogee crest is positioned at the reservoir's normal operating level, equivalent to 1296.5 meters above sea level. The downstream curve of the ogee is defined by the equation $y = 0.1309x^{1.775}$ with a design head of 6.5 meters. This curve transitions into the chute bottom via a 3.5-meter-long circular arc with a radius of 24.31 meters and a central angle of 30.874 degrees. The chute bottom starts at an elevation of 1291.12 meters above sea level, located 33.16 meters downstream of the control section axis.

The concrete chute guides the overflow to the downstream river via a flip bucket. The chute has a rectangular cross-section, starting at 80 meters wide at the upstream end and tapering to 40 meters wide at the downstream end. The chute bed begins at an elevation of 1291.12 meters and maintains a slope of 4.05% over a length of 166.6 meters, descending to 1284.373 meters above sea level. At this point, the slope increases via a convex arc with a radius of 25 meters and a central angle of 15.17 degrees, reaching 30.53%. The convex arc terminates at an elevation of 1283.238 meters, where the chute continues linearly with a 30.53% slope over a length of 111.13 meters, connecting to the flip bucket.

The flip bucket, positioned 295.27 meters downstream of the control section axis, safely disperses flow into the downstream river. The structure maintains a fixed width of 40 meters, with chute and flip bucket sidewalls 3.5 meters high to confine the flow and prevent spillage.

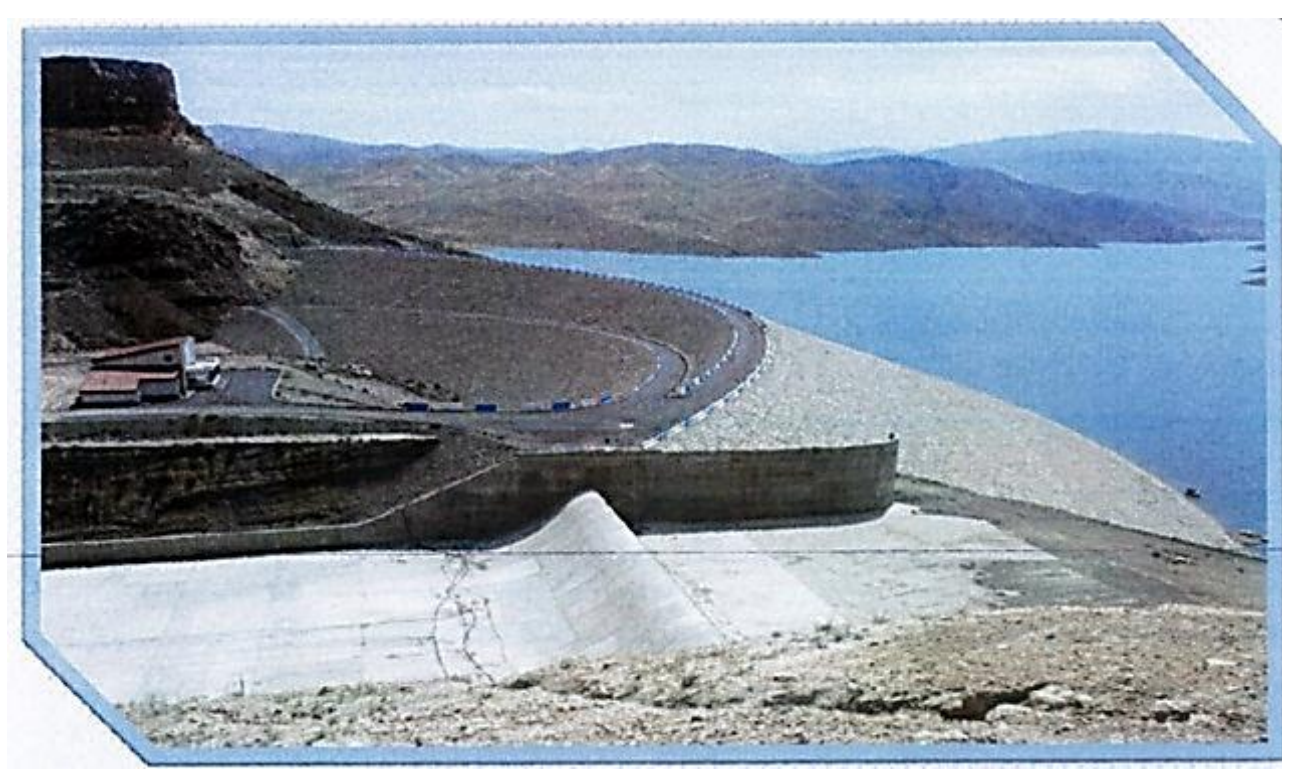

*Figure 2 Aghchai Reservoir Dam*

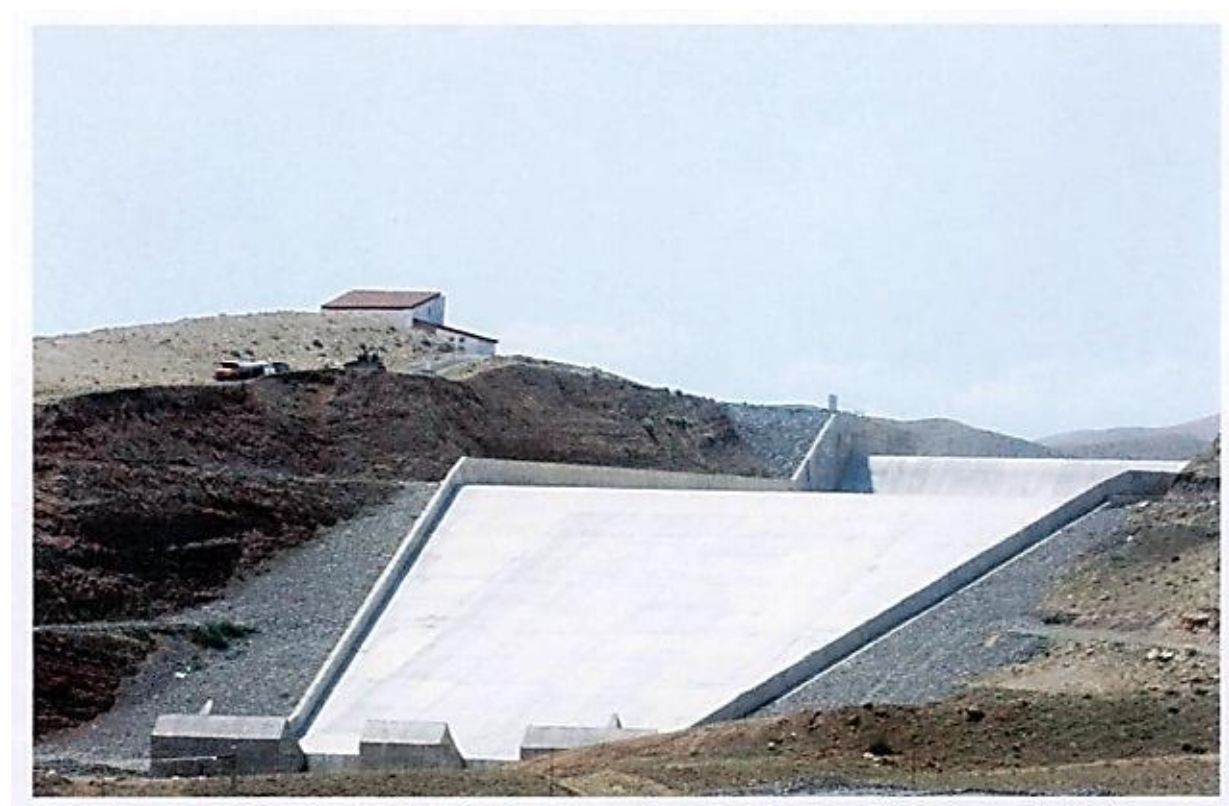

*Figure 3 Ogee spillway of Aghchai dam*

## 2.4. Geometric Modeling and Mesh Generation

The software used for this study is Flow-3D version V10.1 which provides the capability to define meshing and boundary conditions necessary for numerical simulation and sensitivity analysis related to the specified parameters. Additional adjustments were calibrated and simulated to evaluate the influence of hydraulic characteristics on the validation model. The spillway's design accounts for a maximum flood capacity of 4400 cubic meters per second. However, to address the cavitation potential in smaller floods and to ensure alignment with design consultant data, a flow rate of 1065 cubic meters per second was also investigated and modeled. The fluid is treated as incompressible with a free surface, configured within the General Tab of the software. The overall simulation time is set between 60 and 70 seconds, allowing the free surface profile to stabilize.

The numerical model employs the SI system with three core settings: gravity and a non-inertial frame to account for fluid weight, viscosity and turbulence for flow behavior, and cavitation modeling for simulating cavitation phenomena. Gravitational acceleration is specified as $9.81\ m/s^2$ along the y-axis. The fluid is defined as water at 20°C, characterized by Newtonian viscosity. Turbulence is modeled using the RNG turbulence model. For cavitation, the liquid vapor pressure of water at 20°C is specified as 2339 pascals (with atmospheric pressure set to 100 kilopascals), and the characteristic time for vapor bubble collapse is set to microseconds, ensuring accurate representation of cavitation dynamics.

While Flow-3D includes tools to generate simple geometries like cubes and cylinders, the complex profile of the Aghchai Dam spillway required external modeling. The spillway geometry was created using AutoCAD and imported into SolidWorks for refinement. The finalized model was exported as an STL file and then imported into Flow-3D for simulation. Figure 4 illustrates the constructed geometry and its control volume. To optimize computational efficiency and cost, only the right half of the symmetrical spillway, with a width of 40 meters, was modeled.

A critical aspect of accurate numerical modeling involves defining realistic boundary conditions, especially at inlet and outlet boundaries. Figure 4 depicts the defined boundaries for the spillway model. Symmetry conditions were applied along the central plane of the spillway to exploit its geometric symmetry. Rigid wall conditions were applied to the retaining walls and base of the model, ensuring no impact on flow behavior. These conditions are inherently symmetric and do not alter the simulation results. However, specific attention was given to the inlet and outlet boundaries, as they significantly influence flow dynamics.

The inlet boundary conditions were configured to represent the specified overflow discharge rates of 4400 and 1065 cubic meters per second. Due to the symmetry of the model, only half the flow rate was applied, corresponding to 2200 and 532.5 cubic meters per second, respectively. Other input parameters were kept at default settings. Outflow boundary conditions were defined at the model's outlet, allowing fluid to exit freely to prevent accumulation and distortion of the free surface profile. Additionally, the upper free surface was assigned a specified pressure boundary condition with a fluid fraction value of zero, ensuring accurate representation of the free-surface interface.

This refined methodology ensures a high level of accuracy in simulating the hydraulic and cavitation characteristics of the Aghchai Dam spillway, providing reliable insights for validation and analysis.

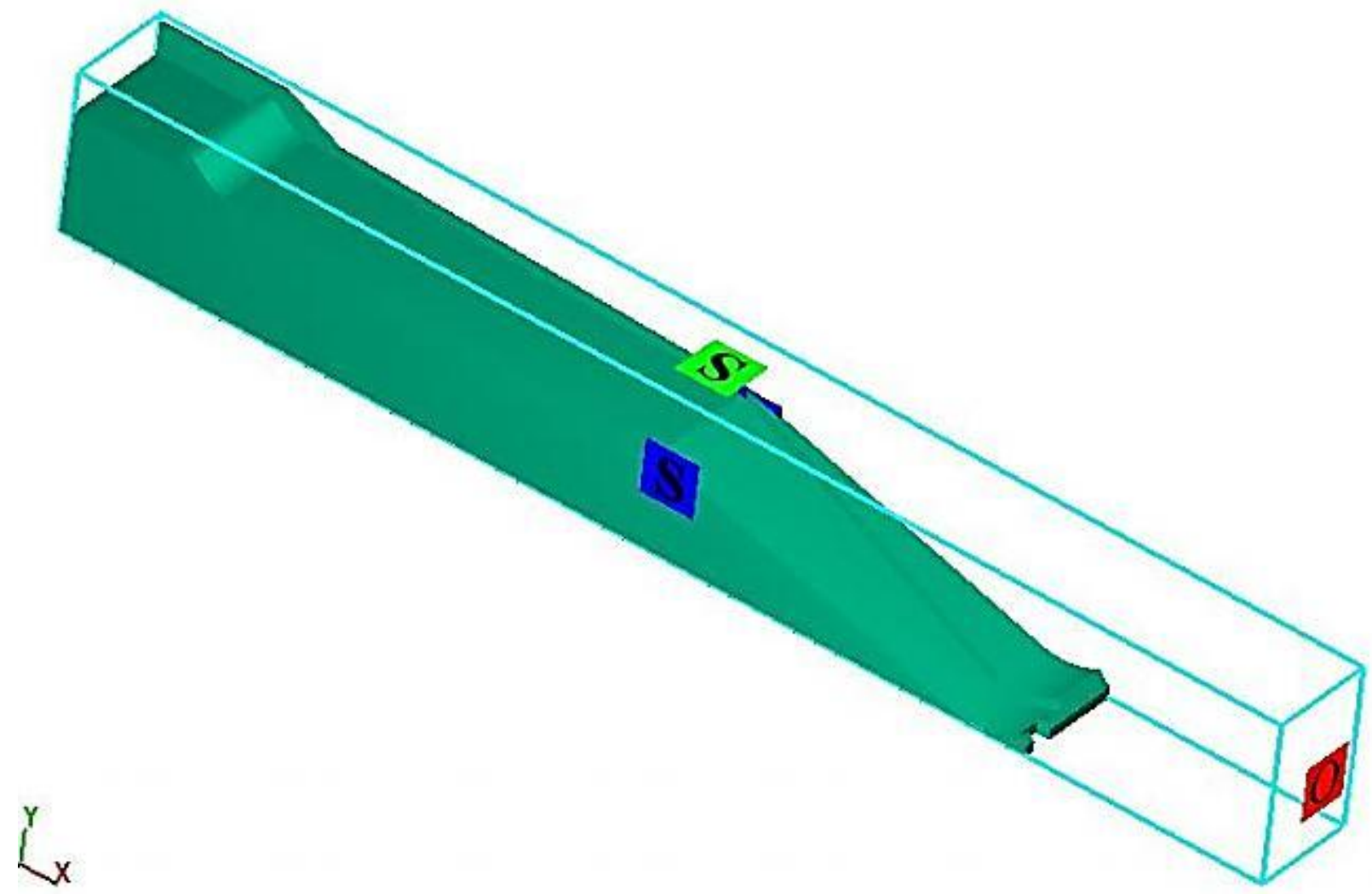

*Figure 4 An example of control volume boundary conditions*

The finite volume solution method on unstructured grids used in this study is primarily based on flux splitting. During the numerical integration of the governing equations, cell-centered methods are applied to solve the flow variables. In Flow-3D, there are two primary approaches for solving the time step in flow equations: implicit and explicit methods. Both approaches play a critical role in determining the stability and accuracy of the simulation.

For this model, the initial time step is set to 0.001 seconds. The software employs an implicit solution method for pressure as the default, which ensures unconditional stability. Unlike explicit methods, which require constraints on the time step size to maintain numerical stability, the implicit method allows for flexibility in time step selection without compromising the stability of the solution. This characteristic makes the implicit approach particularly effective for simulations involving complex flow dynamics.

The flow hydraulic parameters are incorporated into the model outputs and extracted after the simulation is complete. Sensitivity analysis, as illustrated in Figure 5, demonstrates that an explicit method with a time step larger than 0.005 seconds achieves sufficient stability for this specific simulation. These findings highlight the importance of calibrating time step sizes to balance computational efficiency with numerical stability, ensuring the reliability of the simulation results.

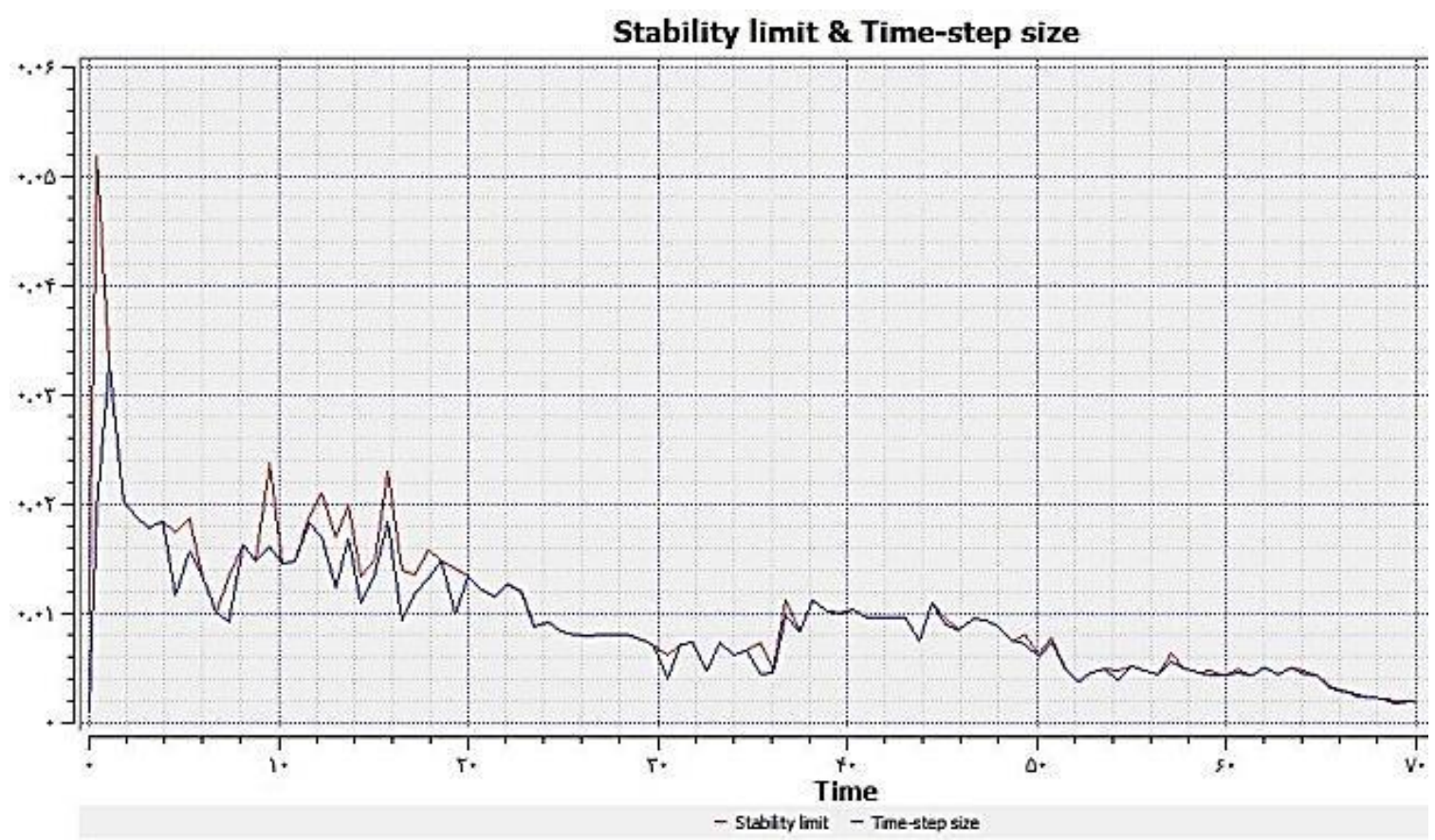

*Figure 5 Time step and stability limit in the method of solving equations*

After completing the modeling process, the model undergoes numerical simulation. Based on the initial conditions outlined in Figure 4, Flow-3D evaluates appropriate time steps, starting with the defined total simulation time of 70 seconds in the General tab. Through a trial-and-error approach, the software refines the time steps to achieve a stable and accurate final state. Similarly, the mesh resolution is optimized through iterative adjustments, and a grid size of 0.5 meters is selected. This resolution ensures that the output flow from the model remains consistent and does not exhibit significant changes when further mesh refinement is applied.

The simulation workflow consists of three key stages: pre-processing, error correction, and the actual simulation. During pre-processing, the model setup is validated, and any errors detected are rectified using tools available in the Display tab, which facilitates the identification and resolution of modeling issues. Throughout the simulation, careful monitoring of intermediate results ensures that the solution remains stable and non-divergent. This iterative approach guarantees the accuracy and reliability of the model.

To optimize computational efficiency, the time interval for simulation steps is progressively increased, provided that no significant variations are observed in the water surface profile during the final stages. These measures ensure a balance between computational speed and the precision of the simulation results, delivering robust and dependable outputs for further analysis.

## 3. Results and Discussions

This section presents the simulation results for various discharge conditions to evaluate the hydraulic performance and cavitation risks of the Aghchai Dam spillway. The analysis includes three-dimensional visualizations, velocity distributions, and cavitation potential contours for flow rates of 4400 and 1065 cubic meters per second.

### 3.1. Discharge of 4400 Cubic Meters per Second

Figure 6 illustrates the three-dimensional simulation results after 70 seconds of runtime, providing a comprehensive visualization of the flow behavior across the spillway. Complementing this, Figure 7 presents a two-dimensional side view of the simulation, showcasing the average velocity distribution along the spillway. These visualizations highlight critical flow characteristics, enabling a detailed assessment of hydraulic performance and cavitation potential within the modeled system.

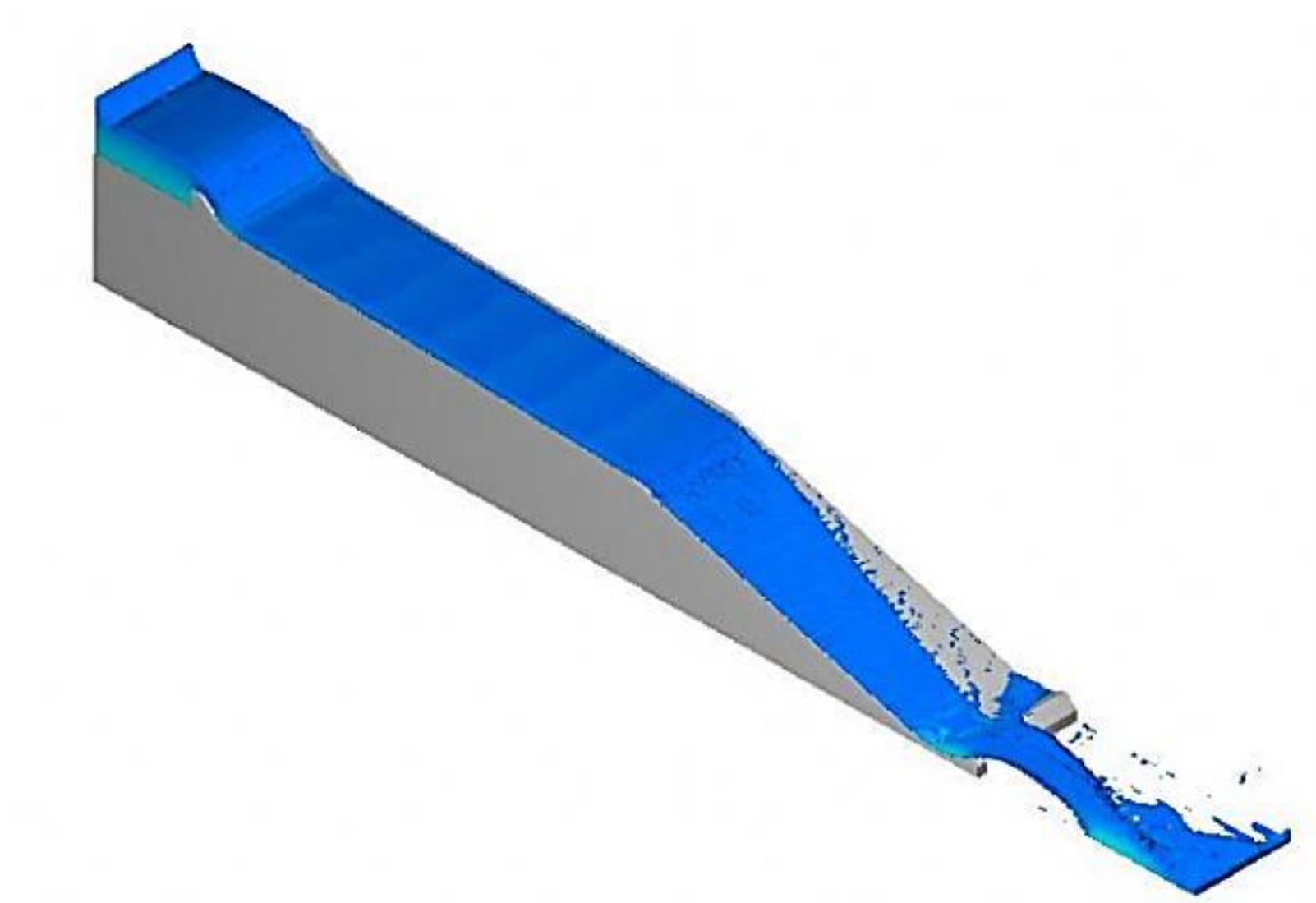

*Figure 6 Maximum flood flow on the spillway in three dimensions at a flow rate of 4400 cubic meters per second.*

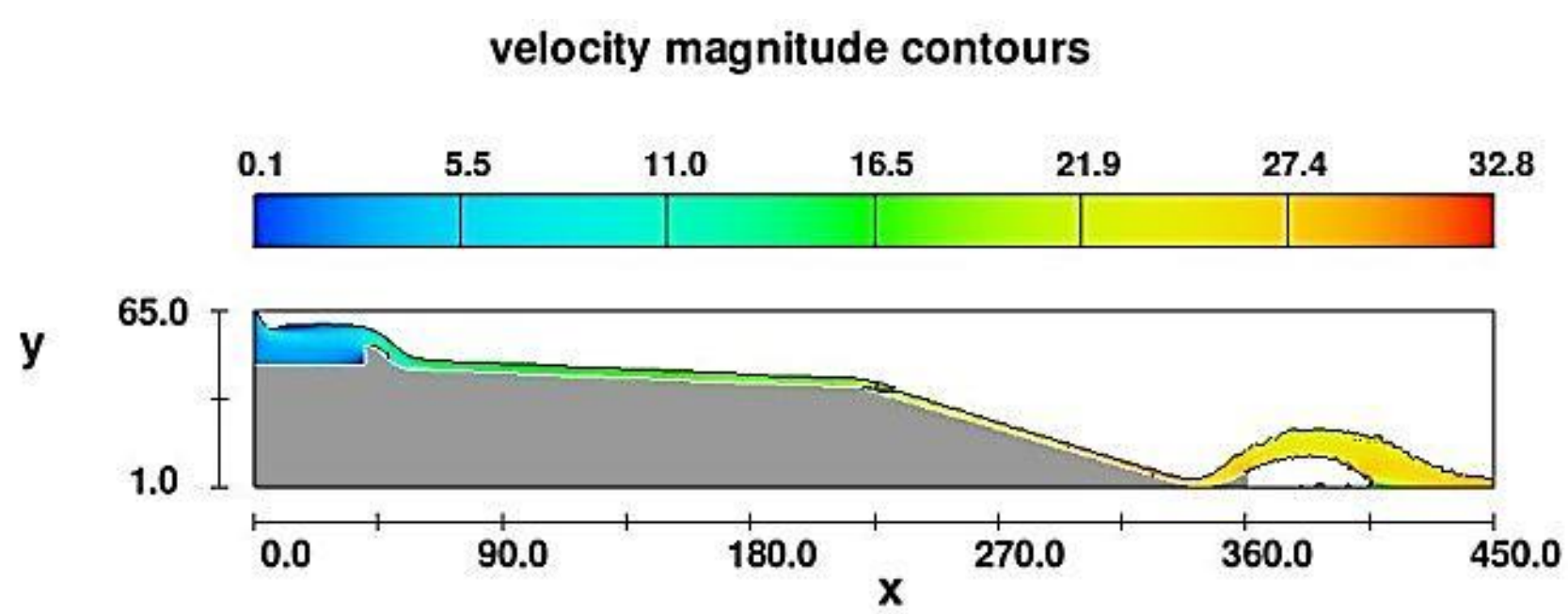

*Figure 7 Contour of the average speed at different points of the overflow in meters per second at a flow rate of 4400 cubic meters per second.*

Flow-3D indicates that the flow achieves a steady state after approximately 52.6 seconds. However, for improved accuracy, the results after 70 seconds have been verified and are presented here. It is important to note that the flow remains unstable before 52.6 seconds, emphasizing the necessity of selecting an appropriate time step based on the analysis in Figure 5. Initially, the water velocity increases as it moves toward the chute’s end, but after stabilization, the velocity decreases and converges to a uniform value, signifying fully stabilized flow conditions.

As the floodwater traverses the spillway, it first flows over the ogee crest and enters the chute, eventually reaching the shooter flip bucket. The flip bucket propels the flow into the air, dissipating its energy before it is deposited into the stilling basin. Observing water level variations and comparing them to the wall height confirms that the spillway is capable of safely handling the design flood of 4400 cubic meters per second, directing all floodwater into the stilling basin without overtopping.

Figure 7 illustrates the velocity profile, showing the lowest velocity in the directing channel before the ogee section and the highest velocity of 32.8 meters per second at the shooter flip bucket. Although this velocity is significant, energy dissipation occurs along the chute due to the considerable height of the overflow and the rapid flow conditions. The flow through the spillway remains in a supercritical state. Figure 8 further demonstrates the turbulence energy profile, with the highest turbulence levels concentrated near the spillway bed and around the shooter flip bucket, reflecting the intense hydraulic activity in these regions.

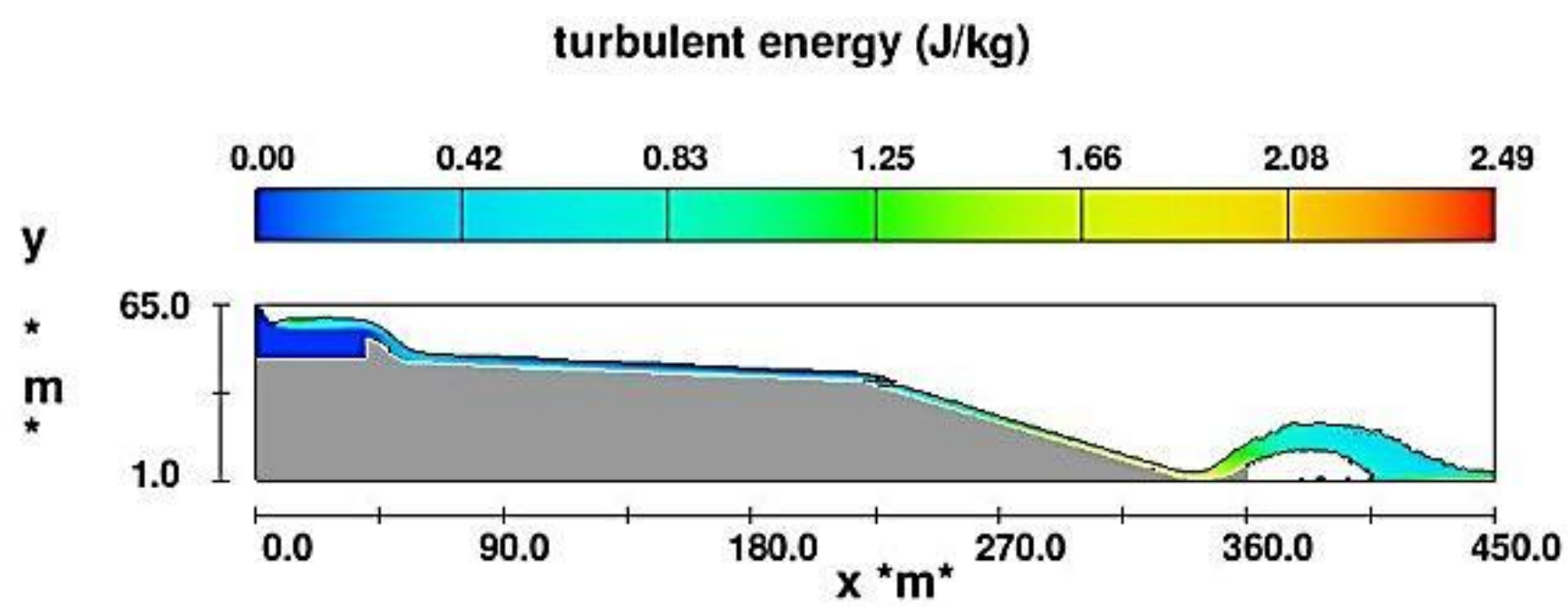

*Figure 8 Turbulence energy contour at different overflow points*

Due to the extremely high fluid velocity (32.8 m/s) observed at high spillways such as the Karun 3 dam spillway, even minor surface irregularities, measuring just a few millimeters, can lead to flow separation. Factors such as abrupt changes in cross-section, surface protrusions, or transitions in the overflow bed geometry—such as those present in the ogee section—can exacerbate this separation. Consequently, the surface of the Aghchai Dam service spillway, situated 65 meters below the peak ogee crest level, is highly susceptible to cavitation.

To assess the risk of cavitation, Figure 9 presents the water pressure contours along various sections of the Aghchai Dam service spillway. These contours provide critical insights into pressure distributions and highlight areas prone to cavitation, guiding the development of mitigation strategies to protect the spillway surface.

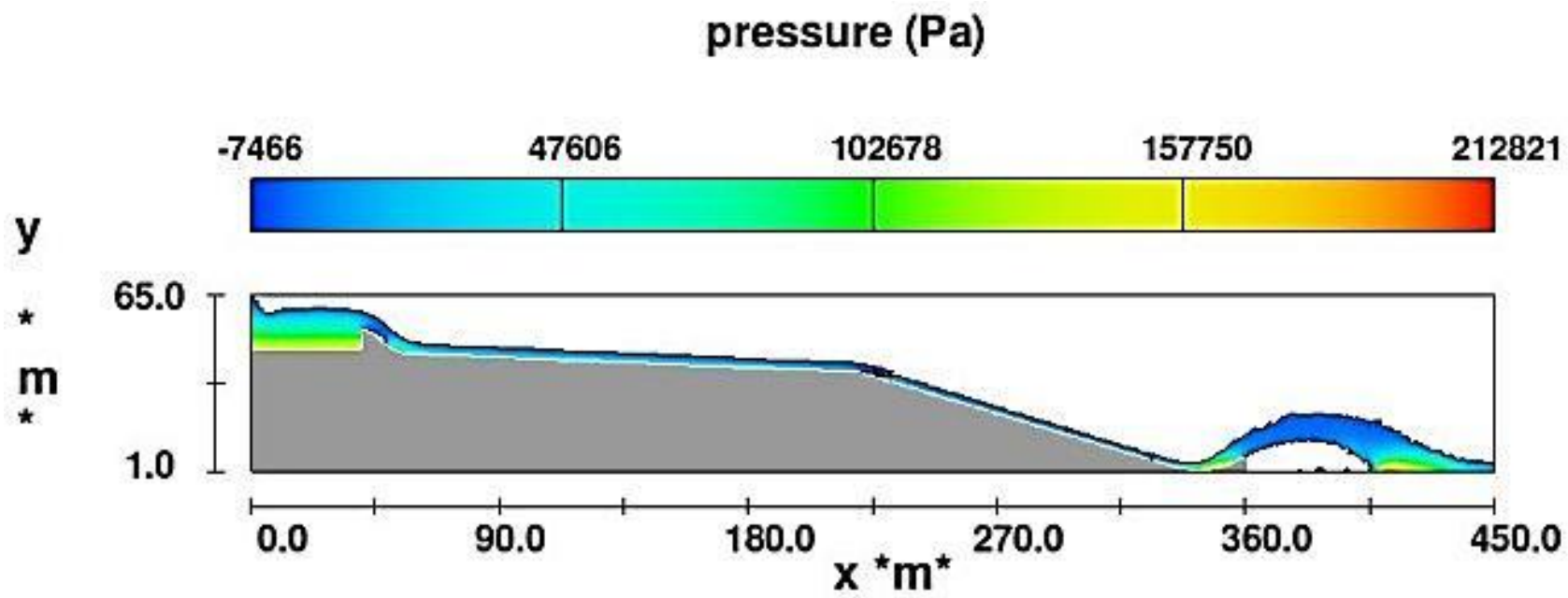

*Figure 9  Pressure contour at different overflow points*

The water pressure across the spillway ranges from a maximum of 212,821 pascals at the flip bucket footer's bed to a minimum relative pressure of 7,466 pascals in various locations. This pressure variation indicates a high potential for cavitation in areas where the pressure drops below the fluid's vapor pressure. For water at 20°C, the vapor pressure is defined as 2,339 pascals, according to the cavitation model used in Flow-3D. The software identifies areas prone to cavitation by capping the pressure meter at 2,339 pascals. In regions where the pressure falls below this threshold, water begins to boil, leading to cavitation. This phenomenon is visually represented in Figure 11, which highlights cavitation potential points on the spillway.

As shown in Figures 10 and 11, cavitation occurs at two critical locations along the overflow: the upper part of the ogee crest and the slope transition in the chute. Cavitation at the crest results from the separation of the flow from the bed at the peak, combined with the high flow velocity and the significant pressure drop. Similarly, cavitation at the chute transition is caused by the abrupt change in slope from 4.5° to 30.5°, leading to localized pressure drops and rapid bubble formation. These bubbles collapse explosively, creating vacuums and increasing the risk of cavitation-induced damage.

Cavitation poses a significant threat to structural integrity when it occurs near walls, as it can cause surface erosion and severe damage over time. Investigations of overflow structures (through both prototype testing and modeling) reveal that cavitation damage results from the interplay of multiple factors, including geometric, hydrodynamic, and material properties. Typically, no single parameter is sufficient to induce cavitation; rather, it is the combination of these factors that results in cavitation damage.

To mitigate cavitation damage, the following strategies can be employed (42)

- Control of the cavitation index for the spillway geometry.
- Flow aeration to reduce cavitation potential.

For the Aghchai Dam service spillway, modifications to the crest geometry are recommended. The peak crest curve, defined by the relationship $y = 0.1309x^{1.775}$can be adjusted by slightly reducing the power exponent to flatten the curve and increasing the width near the origin. This adjustment would extend the crest and reduce the likelihood of flow separation and vacuum formation between the water and the concrete bed. If these changes do not adequately mitigate cavitation, aeration techniques should be employed.

To address cavitation issues at the chute slope transition, the slope of the lower section can be reduced and elongated to minimize abrupt changes, though this approach must consider topographical and economic constraints. Alternatively, a more practical solution involves installing aeration grooves connected to aeration pipes embedded in the surrounding walls. These grooves are designed with a ridge on the floor upstream of the groove, which separates the flow from the bed, allowing effective aeration. Ventilation tubes on the sides introduce air, compensating for pressure deficits and preventing cavitation.

As depicted in Figures 10 and 11, cavitation potential is also observed downstream of the flip bucket, where the pressure falls below the vapor pressure, forming a milky mixture of air and water bubbles. It is critical to note the difference between vapor and air bubbles: vapor bubbles collapse explosively when subjected to high-pressure zones, causing damage, while air bubbles merely cloud the water without posing a risk of explosion. This distinction highlights the importance of aeration in mitigating cavitation damage effectively.

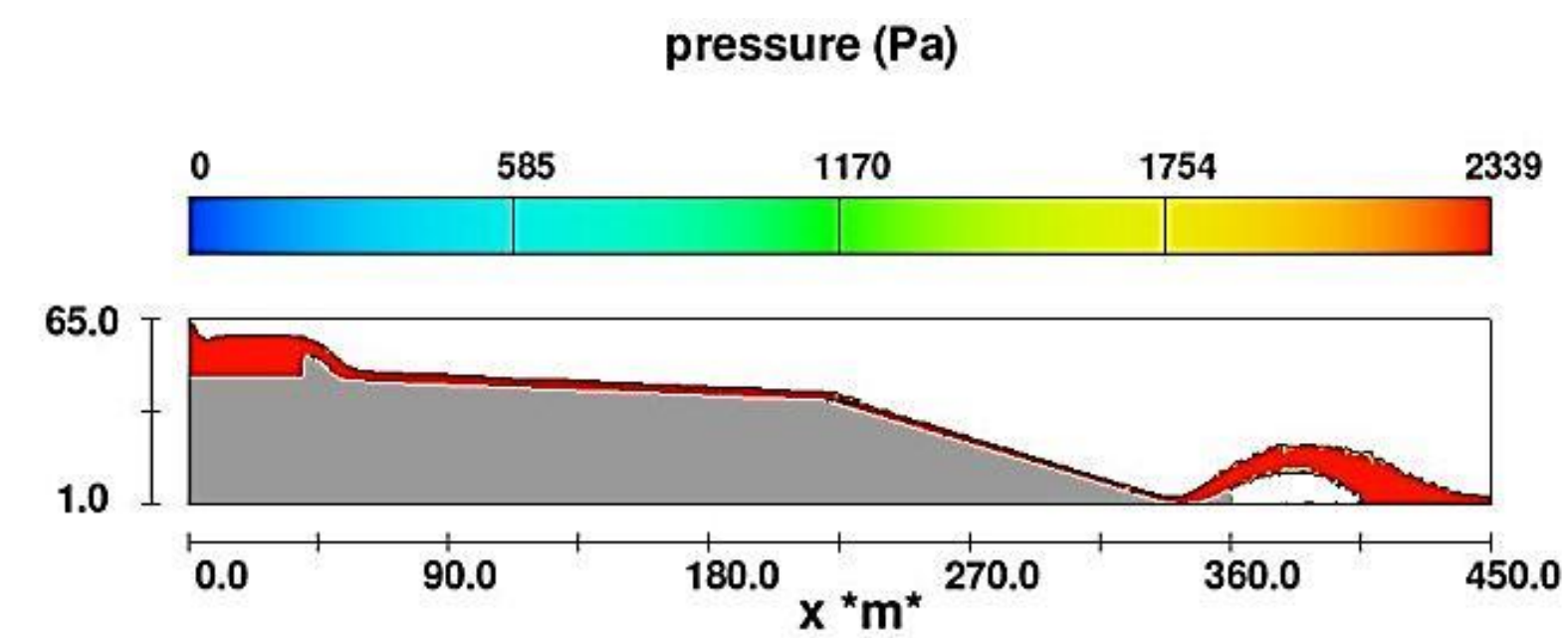

*Figure 10 Pressure contour at different overflow points with a minimum limit of zero and a maximum water vapor pressure of 20 degrees Celsius*

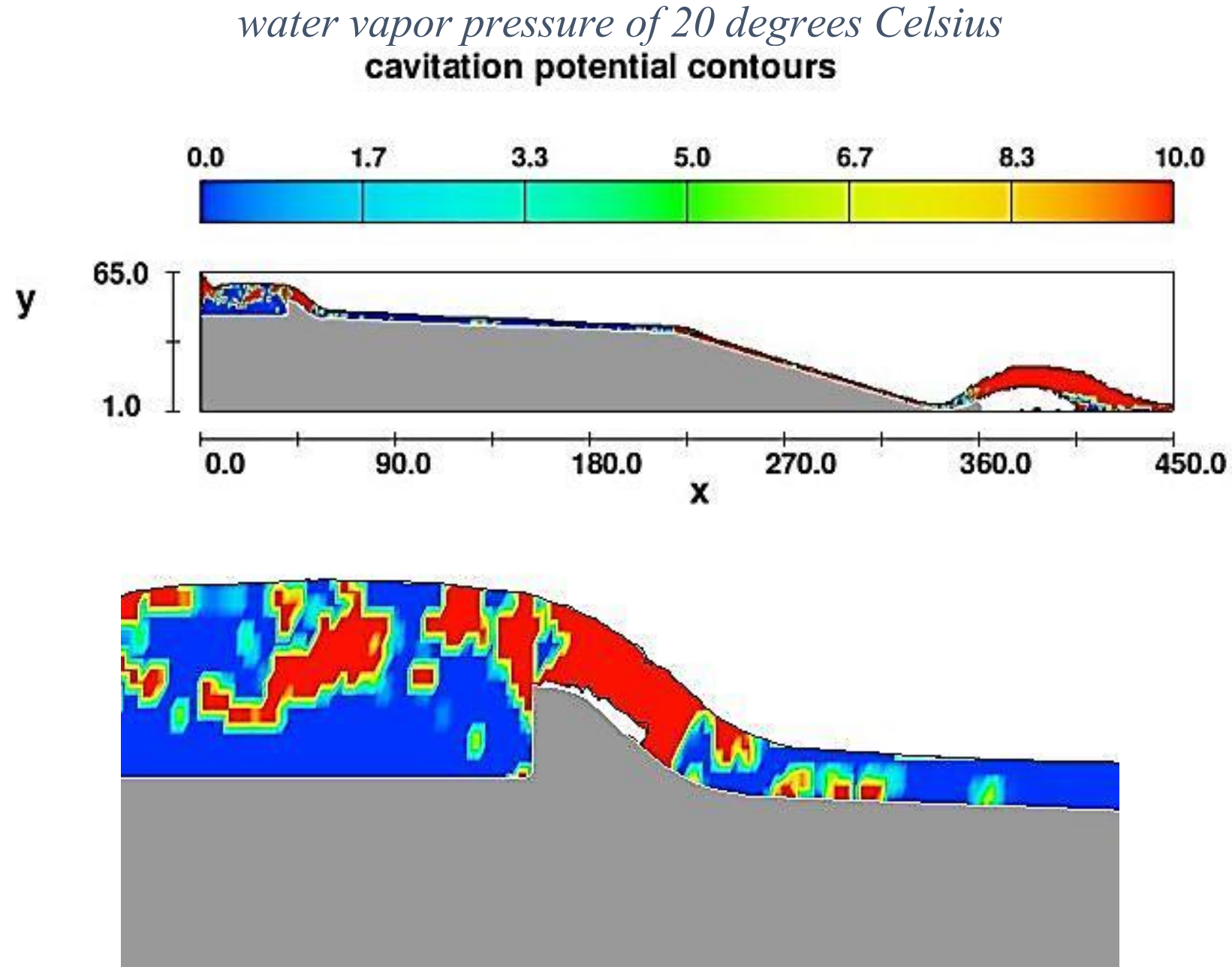

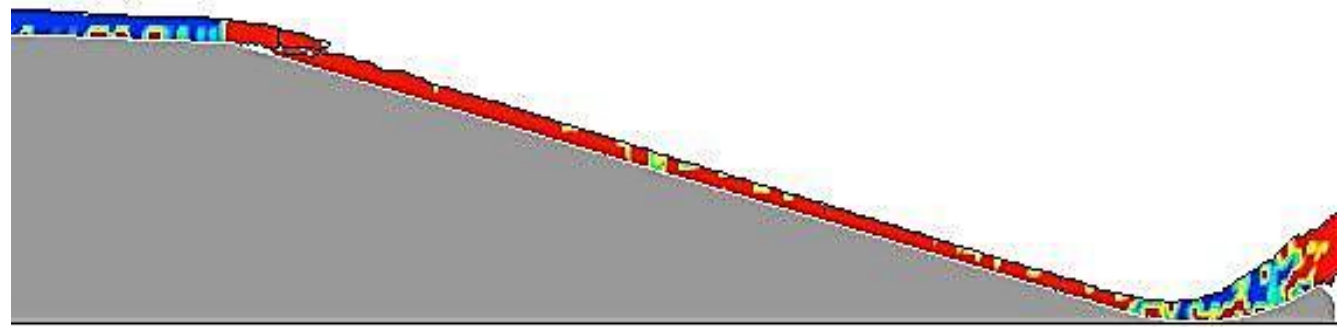

*Figure 11 The contour of points with cavitation potential on the spillway at a flow rate of 4400 cubic meters per second*

## 3.2. Discharge of 1065 Cubic Meters per Second

After executing the simulation and iterating the calculations, it was determined that the flow over the spillway reaches a steady state at 63 seconds for a discharge rate of 1065 cubic meters per second. Figure 12 presents the three-dimensional simulation results, while Figure 13 illustrates the average velocity distribution in two dimensions. Due to the reduced discharge, the maximum velocity observed at the flip bucket footer is 19.7 meters per second. This velocity is intermittent, attributed to the combination of the lower flow rate and the high gradient at the chute's downstream end.

The cavitation potential is visualized in Figure 14, which displays the cavitation potential points contour. The results indicate a possibility of cavitation at both the crest peak and the slope transition of the chute. However, the reduced flow velocity at this discharge rate prevents significant flow separation from the bed, minimizing the likelihood of vacuum formation or cavitation bubble generation. Consequently, the potential for cavitation-induced damage at a flow rate of 1065 cubic meters per second is negligible and does not pose a significant concern.

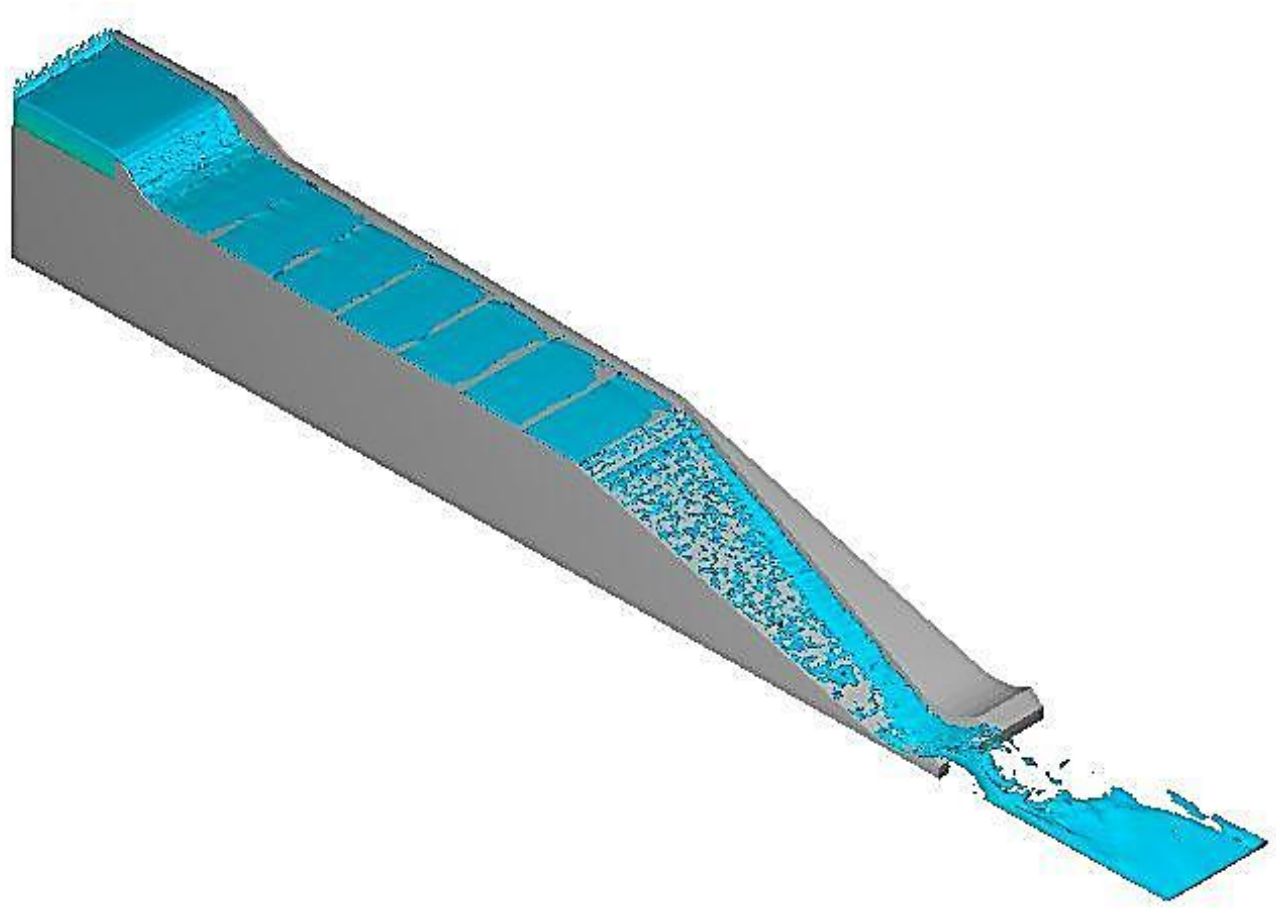

*Figure 12 Three-dimensional view of the maximum flood flow on the spillway at a flow rate of 1065 cubic meters per second*

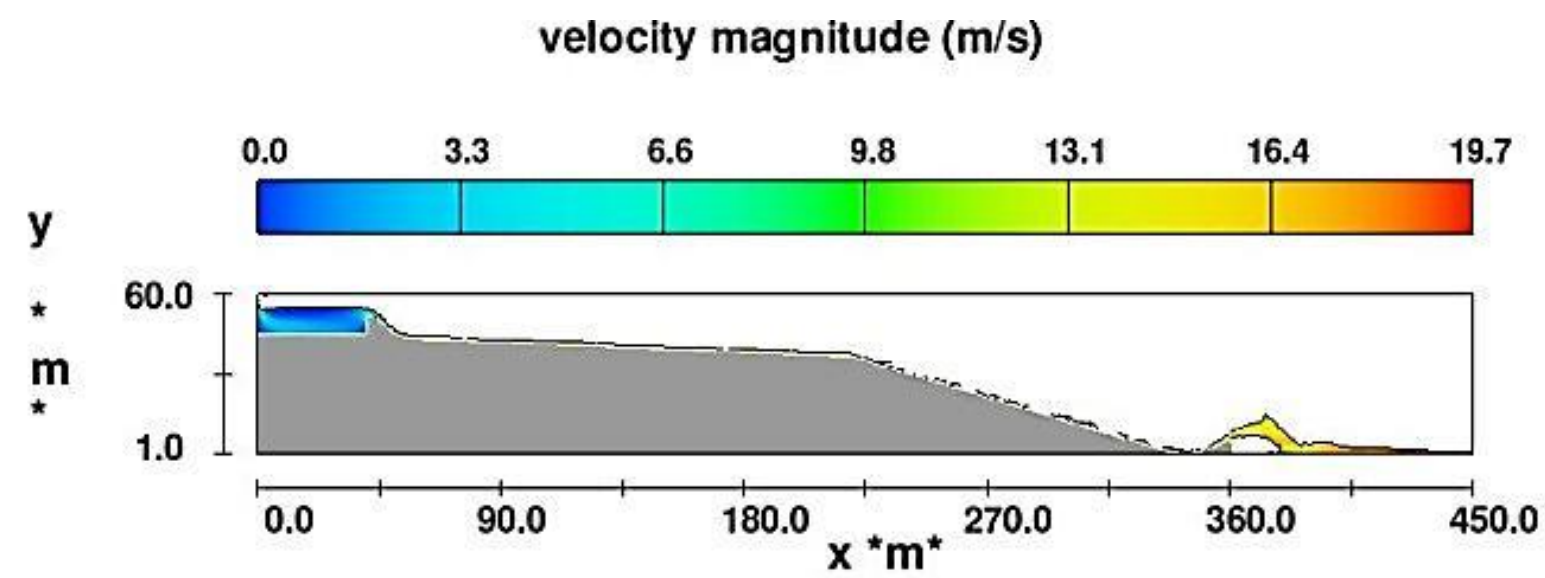

*Figure 13  Contour of the average speed at different points of the overflow in meters per second at a flow rate of 1065 cubic meters per second.*

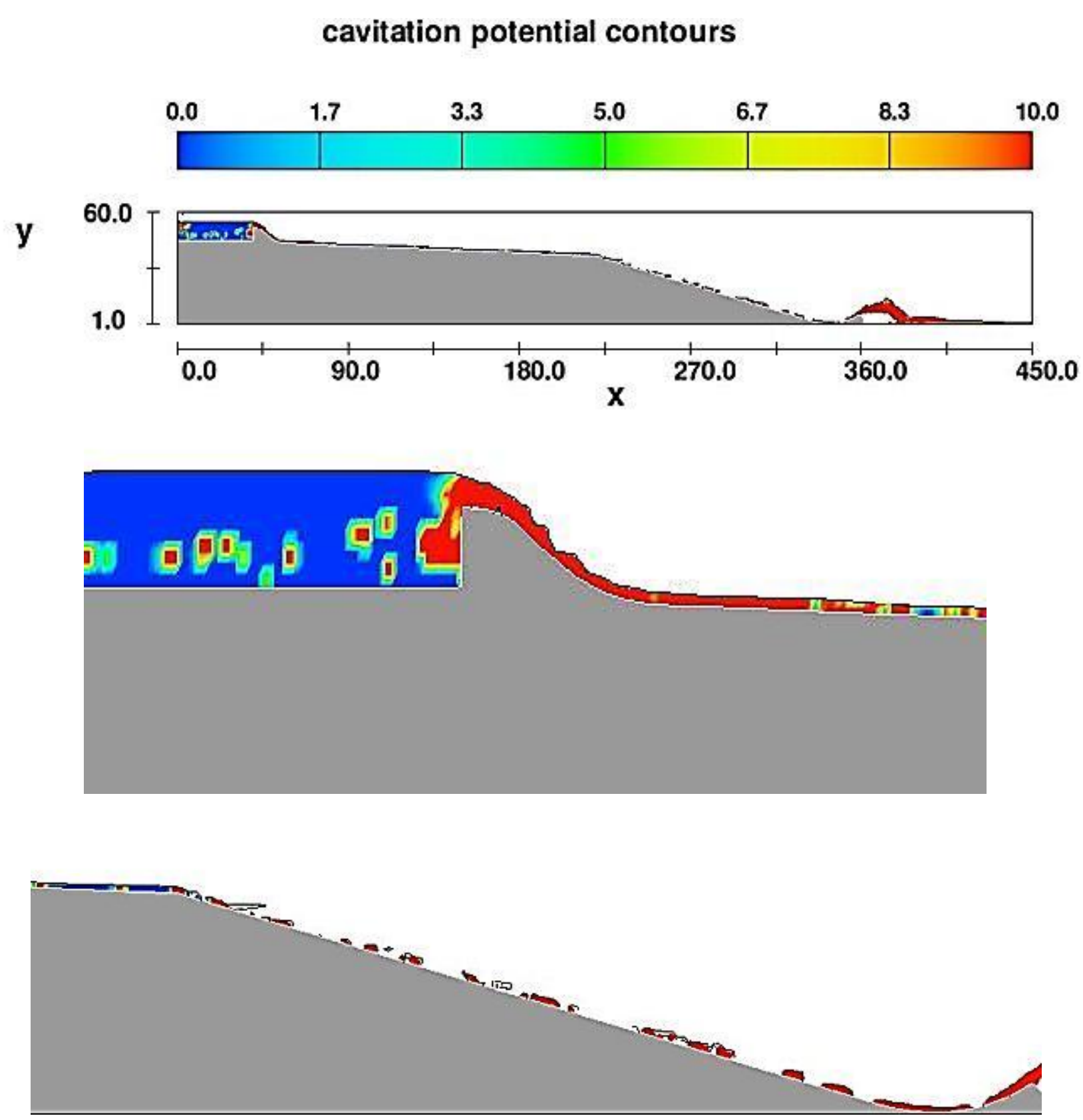

*Figure 14 Contour of potential cavitation points on the spillway at a flow rate of 1065 cubic meters per second*

## 3.3. Mahab-Quds Project

The Mahab-Quds project design consultant (43) utilized the FLD4 computer program to simulate flood propagation dynamics. This specialized software employs the hydraulic pulse method to analyze and evaluate the effects of variations in mountain flood events as they traverse the reservoir. The program provides critical insights into how changes in flood characteristics impact reservoir performance. Figure 15 illustrates the flood propagation results, showing the relationship between reservoir water height and flood discharge.

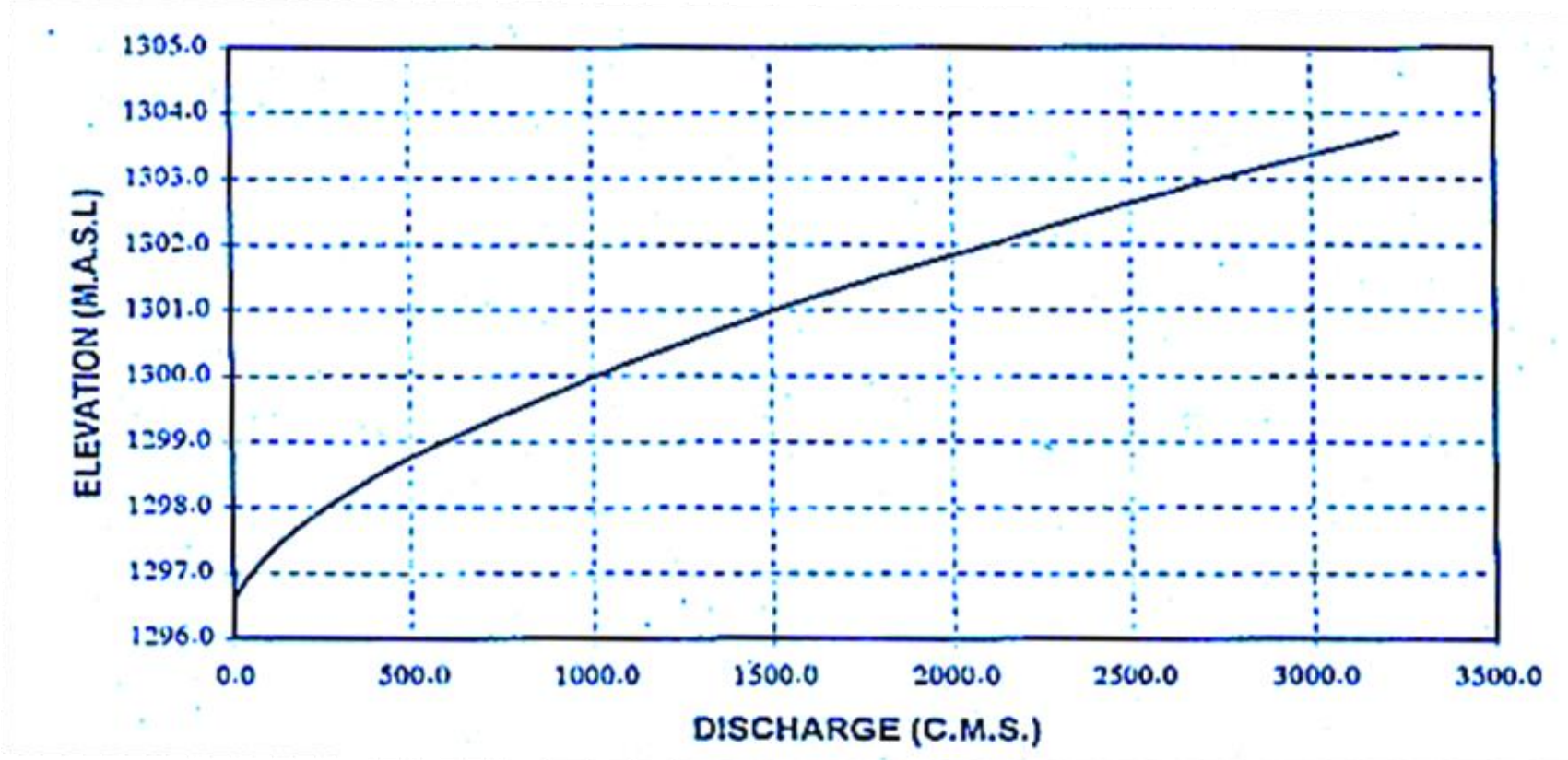

*Figure 15 Reservoir number curve and flood discharge with FLD4 program*

The cavitation condition was evaluated using three key parameters: the risk condition, the cavitation number, and the risk potential under varying surface roughness scenarios. These parameters were calculated for different flow rates—1065, 2253, and 4400 cubic meters per second—using the WS77 computer software developed by the project consultant. Table 1 summarizes the results, providing a comprehensive assessment of cavitation risks for the specified flow conditions.

*Table 1 Hazard status and cavitation number*

| Total Flow Passing Through Spillway ($m^3/s$) | Reservoir Level (m) | Cavitation Risk Condition | Minimum Cavitation Number | Maximum Damage Potential |
|---|---|---|---|---|
| **1065** | 1300.23 | 0.025 | 0.33 | 95 |
| **2253** | 1302.26 | 0.11 | 0.24 | 150 |
| **4400** | 1304.31 | 0.29 | 0.13 | 670 |

According to the WS77 output results, the minimal cavitation damage at a flow rate of 1065 cubic meters per second is negligible, calculated at only 0.025%, and is therefore not considered significant. However, as the flow rate increases to 2253 and 4400 cubic meters per second, the potential for damage escalates substantially. At the maximum design flood discharge of 4400 cubic meters per second, catastrophic damage is possible, which could compromise the structural integrity of the peak section, causing it to lose its original form.

A comparison between the results provided by Mahab Qods consultants and the simulations conducted in FLOW-3D reveals consistent findings. Both analyses indicate that the risk of cavitation rises with an increase in flood flow rate, reaching its peak at the maximum flow rate of 4400 cubic meters per second. The strong correlation between the results of the two methods validates the accuracy and reliability of the findings.

## 4. Conclusion

This study demonstrates the effectiveness of numerical modeling as a reliable and cost-efficient tool for analyzing the hydraulic performance and cavitation risks of the Aghchai Dam service spillway. Using the Flow-3D software and the Volume of Fluid (VOF) method, the research offers valuable insights into the behavior of flow over the spillway under various discharge conditions, replacing the need for expensive and time-intensive laboratory models.

The findings confirm that the Aghchai Dam spillway is capable of safely handling the maximum design flood discharge of 4400 cubic meters per second, effectively transferring the floodwaters to the relaxation basin. Velocity analysis reveals critical flow characteristics, with the highest speed of 32.8 meters per second occurring at the flip bucket footer and significant turbulence localized near the spillway bed and flip bucket areas. These observations underscore the importance of precise spillway bed profile design to ensure flow stability and minimize turbulence-induced damage.

Cavitation was identified at two critical locations: the crest peak and the slope transition. To address these risks, the study recommends adjustments to the ogee crest curve and the slope transition geometry. Extending the cavitation-prone section and reducing slope steepness, where feasible, can effectively mitigate cavitation risks. Alternatively, implementing aeration grooves connected to aeration pipes offers a practical solution for areas where geometric modifications are constrained by topography.

The study validates the applicability of the restricted volume model and the VOF method for modeling free-surface flow and cavitation potential. These techniques align well with the dam design consultant’s results, confirming their reliability for spillway design and analysis. By integrating these numerical tools with traditional hydraulic engineering practices, the study highlights an effective approach to designing robust spillway systems that minimize cavitation risks, optimize hydraulic performance, and reduce maintenance costs.

This research provides actionable recommendations for enhancing spillway design and addressing cavitation risks. It underscores the potential of numerical modeling to inform decision-making in hydraulic engineering, particularly for high-risk structures like spillways, ensuring their long-term operational efficiency and structural integrity.